# Probabilistic Parameter Estimation Using a Gaussian Mixture Density Network: Application to X-ray Reflectivity Data Curve Fitting

Kook Tae Kim and Dong Ryeol Lee[*]

Department of Physics, Soongsil University, 369 Sangdo-ro, Dongjak-gu, Seoul, 06978, Republic of Korea

* email: drlee@ssu.ac.kr

**Abstract**

X-ray reflectivity (XRR) is widely used for thin-film structure analysis, and XRR data analysis involves minimizing the difference between an XRR curve calculated from model parameters describing the thin-film structure. This analysis takes a certain amount of time because it involves many unavoidable iterations. However, the recently introduced artificial neural network (ANN) method can dramatically reduce the analysis time in the case of repeated analyses of similar samples. Here, we demonstrate the analysis of XRR data using a mixture density network (MDN), which enables probabilistic prediction while maintaining the advantages of an ANN. First, under the assumption of a unimodal probability distribution of the output parameter, the trained MDN can estimate the best-fit parameter and, at the same time, estimate the confidence interval (CI) corresponding to the error bar of the best-fit parameter. The CI obtained in this manner is similar to that obtained using the Neumann process, a well-known statistical method. Next, the MDN method provides several possible solutions for each parameter in the case of a multimodal distribution of the output parameters. An unsupervised machine learning method is used to cluster possible parameter sets in order of high probability. Determining the true value by examining the candidates of the parameter sets obtained in this manner can help solve the inherent inverse problem associated with scattering data.

## 1. Introduction

Recently, in the field of X-ray scattering research, machine learning—in particular, experimental data analysis using artificial neural networks (ANNs)—has received extensive attention. For example, an ANN has been successfully used in conjunction with X-ray diffraction to determine the crystal class of crystalline materials (Park *et al*., 2017; Feng *et al*., 2019) and in conjunction with grazing-incidence small-angle X-ray scattering (GISAXS) to classifying the lateral nanostructure of thin films (Liu *et al*., 2019; Ikemoto *et al*., 2020). One advantage of using an ANN is that it can dramatically shorten the analysis time in experiments that are repeatedly performed within the range of already trained variables. For example, it can be applied to time-resolved or in situ experiments or used in sample quality checks performed during mass production processes. A second advantage is that ANN can be useful in areas that are difficult to understand with intuitive human perception. In addition, ANNs can be used to distinguish the characteristics of scattering patterns generated by X-ray scattering, which is difficult using a general regression method.

Among X-ray scattering analysis methods, X-ray reflectivity (XRR) is widely used to obtain the laterally averaged electron depth profile of thin films of various materials at the subnanometer scale (Daillant & Gibaud, 2009). Quantitative analysis of these XRR data is carried out by comparing dynamical calculations performed using, for example, Parratt's recursion formula (Parratt, 1954) or an optical matrix (Daillant & Gibaud, 2009), with experimental data and then obtaining structural parameters such as the density, thickness, and interfacial roughness of each layer inside a thin film. To attain the fitting parameter that best describes the experimental data, the $\chi^2$ minimization method using a nonlinear optimization algorithm is usually used. However, studies that use machine learning technology as a parameter estimation method for XRR data are becoming increasingly common (Mironov *et al*., 2021; Greco *et al*., 2021; Carmona-Loaiza & Raza, 2021). Strauß *et al.* (2004) have shown that the structure of a thin film can be refined after the calculated XRR data have been trained within the range of the pre-specified parameter space using a support vector machine. Greco *et al.* (2019) introduced fast fitting using an ANN for XRR data recorded for an organic thin film under in situ growth. Parameter estimation using an ANN can be carried out at high speed even with an ANN architecture consisting of a relatively small number of nodes.

Although the body of research involving ANNs is increasing in the overall field of X-ray scattering, including in XRR, the literature contains few studies on the reliability of parameter estimation using ANNs. Only recently, studies on prediction uncertainty of neural networks and methods for increasing prediction accuracy have been conducted (Mironov *et al*., 2021; Greco *et al*., 2021). In addition, fundamentally solving the phase problem of X-ray scattering is difficult even if an ANN is used (Carmona-Loaiza & Raza, 2021). That is, because phase information is lost when observing scattering intensity, a one-to-one correspondence between the scattering intensity and structural parameters might not be established and the solution might not be unique. Because ANNs are suitable in cases of one-to-one correspondence between input and output, ANNs may not be appropriate for the inverse problem of finding the structural parameters of a sample from scattering intensity data. Therefore, given this inverse problem, questions arise about the reliability of parameter estimations obtained as a result of XRR curve fitting using a neural network.

In conventional nonlinear optimization problems, the reliability of parameters obtained through curve fitting is usually expressed as a confidence interval (CI) corresponding to the range of parameters for the true parameter to appear as a probability of a certain value around the best-fit parameter. Knowing the error range of the parameter is helpful in determining an appropriate physical model among nonunique solutions; thus, knowing the error range might be important for solving the aforementioned inverse problem. For example, a parameter with an excessively large error range indicates that the model may be overly complex or that a high correlation exists between different parameters.

To solve this problem, we here apply the mixture density network (MDN) (Bishop, 1994)—a neural network based on probability theory—to XRR analysis. MDN has a structure similar to that of an existing neural network known as the fully connected perceptron (FCP) and also enables the probability distribution of parameters to be quickly estimated on the basis of data. MDNs are suitable for inverse problems, in particular, because they provide a possible multimodal distribution of parameters; MDNs have been successfully applied to inverse design for optical transmission (Unni & Zheng, 2020). We show here that applying MDN to XRR data analysis enables a good solution to the inverse problem of scattering to be obtained and the error range of the parameter to be determined.

## 2. Neural network for X-ray reflectivity analysis

### 2.1 X-ray reflectivity

When X-rays are incident on a thin film, absorption and refraction of the X-rays inside the film are described using the real ($\delta$) and imaginary ($\beta$) parts of the refractive index $n$:

$$n = 1 - \delta + i\beta. \quad (1)$$

If the refractive index of the thin-film structure sample is known, the intensity of X-rays specularly reflected by the film can be calculated. To quantitatively describe the reflected X-ray intensity from a nanoscale thin film, the depth profile of the refractive index inside the film must be known. The depth profile of these refractive indexes is often parameterized by the thickness, the interfacial roughness, and the density relative to the bulk value for each layer. The XRR curve is completely determined from these parameters. Let $\boldsymbol{x} = \{x_i\}$ ($i = 1,2,3...k$) be the vector of parameters to explain the structure of the film. The XRR curve can be calculated from the parameter vector $\boldsymbol{x}$ of the sample using a well-known method such as the Parratt recursive formula (Parratt, 1954). This XRR curve is calculated as a function of the incident angle $\theta_l$ ($l = 1,2...M$) and can be described as $R_l = R(\theta_l; \boldsymbol{x})$. After an appropriate model of the thin film is determined, let $N$ parameter sets randomly selected from the range of interest or physically possible for each parameter be $\{\boldsymbol{x}^j\}$ ($j = 1,2...N$). Let the corresponding set of $R_l$ be $\{R_l^j\}$. Our goal is to (1) train the neural network on the relation $\{R_l^j\} \to \{\boldsymbol{x}^j\}$, (2) put the XRR data $R_{\text{obs}}$ obtained through the experiment as input to the neural network, and (3) obtain the best-fit parameter vector $\boldsymbol{x}_{\text{obs}}$ and the error $\boldsymbol{\sigma}_{\text{obs}}$ of the parameter as output in the neural network.

## 2.2 Mixture density network

An MDN enables probabilistic prediction by obtaining the conditional probability density $p(x^j|R_l^j, \mathbf{w})$ of the output for a given input (Bishop, 1994; Unni & Zheng, 2020; McLachlan *et al.*, 1988). Unlike a general FCP, which gives only one best-fit parameter set for a given input as an output, MDN gives values expressing the probability density distribution of parameters as an output. Therefore, MDN can directly obtain the error range or CI of each parameter from the probability distributions of the parameters provided as outputs.

Any probability distribution can be expressed as a linear combination of several Gaussians (McLachlan *et al.*, 1988). Therefore, the probability distribution of parameter $x^j$ can be expressed as a mixture probability consisting of $N_m$ Gaussians. The training process of an MDN is the same as finding the network weight $\mathbf{w}$ that minimizes the following loss function $\mathcal{L}$ by the maximum likelihood method (Bishop, 1994):

$$\mathcal{L} = \sum_{j=1}^{N} E_j, \quad (2)$$

where $E_j$ represents the extent to which the *j*-th data set among *N* randomly selected datasets contributes to the loss function. In the case of MDN, finding the $\mathbf{w}$ value that minimizes the above loss function corresponds to finding the $\mathbf{w}$ value that maximizes the conditional probability density $p(\{x^j\}|\{R_l^j\}, \mathbf{w})$, which is likelihood, for all data sets $\{x^j, R_l^j\}$ used for network training. It can be expressed as a Gaussian mixture model:

$$E_j = -\ln\left(\sum_{m=1}^{N_m} \alpha^m(R_l^j, \mathbf{w}) \phi_m(x^j|R_l^j, \mathbf{w})\right), \quad (3)$$

$$\phi_m(x^j|R_l^j, \mathbf{w}) = \frac{1}{(2\pi)^{\frac{k}{2}} \prod_{i=1}^{k} \sigma_i^m(R_l^j, \mathbf{w})} \exp\left\{-\frac{1}{2}\sum_{i=1}^{k}\left(\frac{x_i^j - \mu_i^m}{\sigma_i^m(R_l^j, \mathbf{w})}\right)^2\right\}, \quad (4)$$

where $\alpha^m$, $\phi_m(x^j|R_l^j, \mathbf{w})$, $\mu_i^m$, and $\sigma_i^m$ represent the mixing coefficient, conditional density, mean, and the standard deviation of the *m*-th Gaussian, respectively. The MDN model gives the mixing coefficient $\alpha^m$, mean $\mu_i^m$, and standard deviation $\sigma_i^m$, which describe the probability density of the output parameter for a given XRR curve, as the network output from the last output layer (Fig. 1). In this case, the most commonly used FCP or convolutional neural network can be entered in the remaining hidden layers except for the output layer.

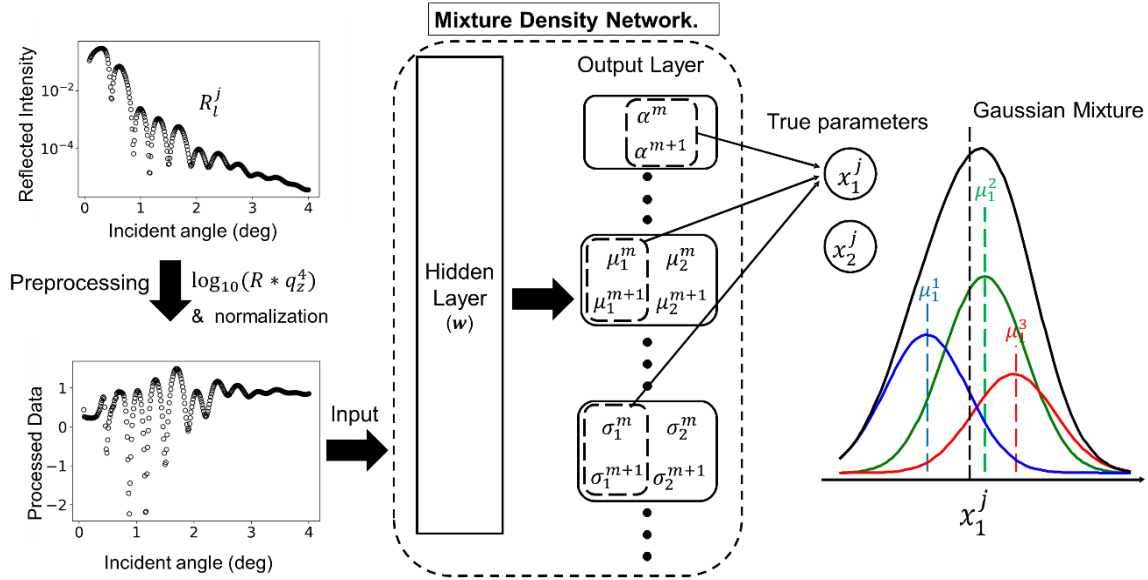

**Figure 1** Schematic of the mixture density network (MDN) for true parameter $x_i^j$. The XRR curve $R_l^j$ becomes the input data after preprocessing. The MDN is composed of hidden layers and the output layer. Hidden layers consist of fully connected perceptrons with weight $w$. The output layer, expressed by the mixing coefficient (α), Gaussian mean (μ), and standard deviation (σ), describes the probability distribution of the true parameter, as shown in the right figure.

The argument of the ln function in Equation (3) must be normalized because it is equal to the conditional probability density at which $x^j$ will be obtained for a given $(R_l^j, \mathbf{w})$. This normalization acts as a constraint on $\alpha^m$, and this constraint is solved by applying *softmax* activation, a normalized exponential function, to all nodes that give $\alpha^m$ as an output. The positive standard deviation $\sigma_i^m$ uses exponential activation, and the unbounded average value $\mu_i^m$ uses linear activation (Bishop, 1994). As a special case of the MDN model, if the output is modeled by only one Gaussian and the standard deviations of all parameters are the same, then the minimization of the loss function in Eq. (2) becomes the same as the well-known minimization of the mean squared error (Bishop, 1994). In this case, the practically important output among the network outputs is only the mean value $\mu_i^m$ and training the MDN is the same as training the regression model using a general FCP.

### 2.3 Data processing and network architecture

For the application of an MDN to XRR analysis, the XRR data for a Pt/Co/Pt trilayer thin film used in a previous study were chosen (Kim *et al.*, 2021). As previously described, the depth profile of the refractive index for the Pt/Co/Pt trilayer can be expressed using parameters such as film thickness, density relative to that of the ideal bulk material, and the interface roughness (Figure 2). The XRR curve was calculated using Parratt's recursive formula (Parratt, 1954) after the depth profile obtained from the parameters was replaced with many slabs of uniform thickness smaller than $2\pi/q_{z,max}$ where $q_{z,max}$ is the maximum $q_z$ value of the experimental data. To create training data, a sufficiently large number of parameter sets were randomly chosen within the predefined range of interest for each parameter and the corresponding XRR curves were calculated. In practice, as a result of training a single-Gaussian MDN model while increasing parameter sets from 10,000 to 1,000,000, the minimum size of 500,000 at which the CI does not change any more, was used in this study. The refractive index of the bulk material was calculated using a well-known reference (Henke *et al.*, 1993). The electron density of each layer was obtained from the real value of the refractive index, and the relative ratio to the value of the bulk material was in the range from 0.7 to 1.1. The thickness of each layer was between 20 and 40 Å, and the interfacial roughness was set between 0 and 8 Å. The ranges of these parameters were empirically determined based on nominal values, and a wider range was selected than when manually fitting.

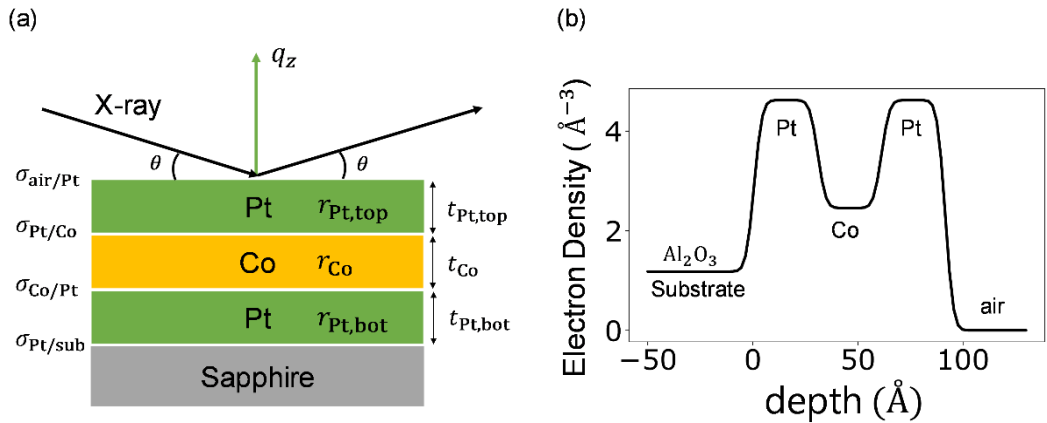

**Figure 2** (a) Schematic of X-ray reflectivity for a Pt/Co/Pt trilayer. The structural parameters *r*, *t*, and *σ* represent the density relative to that of the bulk material, layer thickness, and the interface roughness, respectively. (b) Depth profile of electron density calculated using the structural parameters.

For efficient neural network training, all input data values should have a similar order. However, because the XRR intensity decreases rapidly as the wave vector transfer $q_z$ proportional to the incident angle $\theta$ increases, the calculated XRR intensity is multiplied by $q_z^4$ and the $\log_{10}()$ function is then performed so that the magnitude of the XRR intensity is similar in the entire $q_z$ region. Because the appropriate input data of the neural network must have an absolute value between 0 and

1, it is finally used as input data after scaling using the mean and standard deviation values of $\log_{10}(\text{intensity} \times q_z^4)$ values for all test data. True parameter data corresponding to the output were normalized between 0 and 1 using the maximum and minimum values within the range of interest for each parameter. After data creation, 50,000 data sets, or 10% of the total data, were used as a test dataset to check the training quality after training. The network model used Tensorflow (Abadi *et al*., 2016), which is a Python library, and the network architecture used the structure of the input layer, the hidden layer consisting of 400, 800, 400, and 300 perceptrons, and finally the output layer, where the MDN was used. All activations of the hidden layer were performed using the "relu" function, which gives 0 for input values less than 0 and the same value as input for values greater than 0. To apply the MDN, the mixture normal layer provided by the Tensorflow Probability library (Dillon *et al*., 2017) was used for the final output layer. For $N_m$ mixtures and $k$ output parameters, the output layer contained a total of $(2k + 1)N_m$ output nodes (Figure 1). The program codes used in this paper can be found at https://github.com/KookTaeKim/XRR-MDN.

## 3. Results and discussion

### 3.1 Single-Gaussian density network and confidence interval

We first attempted to investigate the CI or error range of each parameter in the case of estimating the parameter with the commonly used FCP ANN. However, the training of this FCP ANN is almost identical to the training of the density network assuming that the distribution of each output parameter is expressed through a single Gaussian. Therefore, instead of an FCP ANN, which does not provide information about parameter error by itself, a single-Gaussian model MDN was used. To this end, as previously described, after the data processing and network architecture setup, training was performed using the density network with single-Gaussian unimodal output distribution. After training was completed, the mean and standard deviation values of the output parameter vector were obtained as the MDN output for the experimental data as input (Table 1). The mean value corresponds to the best-fit parameter. To determine whether the solution obtained through MDN is an appropriate solution, we calculated the XRR curve using the best-fit parameter and compared it with the experimental data (Fig. 3(a)). At low angles of incidence, the XRR curves agreed well with the experimental data; by contrast, at high angles of incidence, they did not explain the total thickness fringe well.

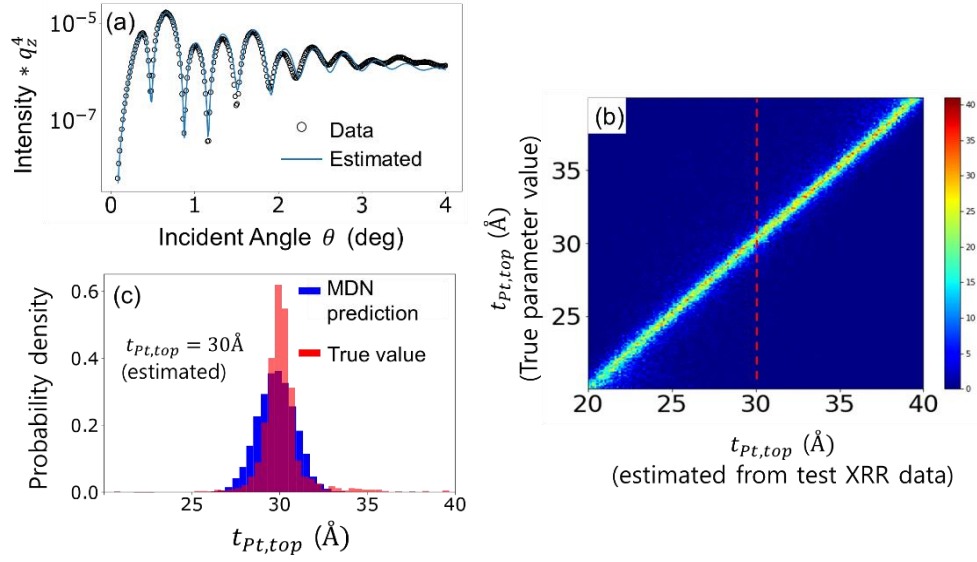

**Figure 3** (a) The X-ray reflectivity (XRR) curve and experimental data calculated using the best-fit parameters obtained from the single-Gaussian density network. (b) Contour plots for the best-fit parameters obtained using single-Gaussian MDN from 50,000 test XRR curve data, and the true parameters used to create each test datum. The red dotted line represents the best fit value of the thickness of the top Pt layer predicted for the experimental data. (c) Probability density function of the top Pt thickness predicted for experimental data from single-Gaussian MDN (blue histogram) and the probability distribution of true parameters (red histogram) along the best-fit value indicated by the red dotted line in (b).

**Table 1** Mean and standard deviation values of the predicted probability distribution for the output parameter obtained using the density network assuming a single-Gaussian output probability density function (PDF). Parameter $r$ is the relative density of the bulk material, $t$ is the layer thickness, and $\sigma$ is the interface roughness.

|  | $r_{Pt,top}$ | $r_{Co}$ | $r_{Pt,bot}$ | $t_{Pt,top}$ (Å) | $t_{Co}$ (Å) |
|---|---|---|---|---|---|
| Estimated | $0.82 \pm 0.05$ | $0.93 \pm 0.02$ | $0.99 \pm 0.04$ | $30 \pm 1$ | $26.1 \pm 0.5$ |

|  | $t_{Pt,bot}$ (Å) | $\sigma_{air/Pt}$ (Å) | $\sigma_{Pt/Co}$ (Å) | $\sigma_{Co/Pt}$ (Å) | $\sigma_{Pt/sub}$ (Å) |
|---|---|---|---|---|---|
| Estimated | $31.3 \pm 0.8$ | $3.6 \pm 0.7$ | $4.5 \pm 0.8$ | $4.8 \pm 0.6$ | $0.7 \pm 0.6$ |

Two approaches can be considered for improving the analysis results. First, XRR is relatively more sensitive to the microstructure in the thin film at high incidence angles than at low incidence angles; thus, an additional interlayer structure likely exists. Since increasing the number of parameters may cause an overfitting problem, more complex structural models will not be discussed further in this study. Second, our intuition is that the curve fitting should be improved over the entire range of incidence angles by compromising the quality of curve fitting at low angles of incidence. However, this artificial control to obtain a physically more meaningful solution is not allowed in the general regression method. Regression methods including a single-Gaussian model network have a fundamental limitation of simply minimizing the difference between the experimental and calculated values and returning a single solution. Considering the multiple solutions due to the inverse problem in X-ray scattering, the global minimum may not represent the actual sample. For example, in the X-ray reflectivity analysis for an A/B multilayer, it is well known that even if the roughness of the A-on-B and B-on-A interfaces are reversed, it is difficult to distinguish them. In this case, classical regression methods or standard neural networks that present us with only one solution have obvious limitations. However, the MDN allowing many-Gaussian mixture probabilities, which will be described later, has the advantage of enabling appropriate selection because it presents a multimodal probability distribution that potentially includes several possible solutions.

We first discuss the error or standard deviation of the output parameter provided by the MDN method involving a single-Gaussian model. In the present study, the dataset used when training the MDN is a well-defined mathematical model's true parameter and true XRR curve. With this approach, when the MDN is trained with the input data of the true model, the standard deviation obtained from the output represents the CI indicating the range of the probability that the true parameter appears around the best-fit parameter. To confirm this relationship between the standard deviation and the CI, first, as shown in Figure 3(b), the estimated parameter obtained for the top Pt layer thickness and the corresponding true parameter distribution were drawn using the test data left unused for training as the MDN input. Second, using the well-known Neumann process for the contour drawn in this manner, the CI for the observed parameters was estimated (Behnke *et al.*, 2013; Cowan, 1998).

The Neumann process is briefly described as follows. (1) When the distribution of the true parameter versus the estimated parameter is known, the conditional density $g(x_{\text{True}}|\hat{x})$ of the true parameter for the estimated parameter $\hat{x}$ can be known. (2) From this conditional density, the probability that the true parameter belongs within CI $[x_{\text{True,up}}, x_{\text{True,low}}]$ for $\hat{x}_{\text{obs}}$ observed through the experiment becomes $(1 - \alpha)$. The latter can be obtained using the following integral equation:

$$1 - \alpha = \int_{x_{\text{True,low}}}^{x_{\text{True,up}}} g(x_{\text{True}}|\hat{x}_{\text{obs}}) \mathrm{d}x_{\text{True}}, \quad (5)$$

where $g(x_{\text{True}}|\hat{x}_{\text{obs}})$ can be obtained by normalizing the projection of the true parameter obtained along the vertical line located at the value of the estimated parameter $\hat{x}_{\text{obs}}$ in the contour plot of the true vs estimated parameter (Fig. 3(b)). The probability density function (PDF) obtained in this manner corresponds to the red histogram in Fig. 3(c). To show that the range of the output parameter of the MDN is equivalent to the CI, we now compare $g(x_{\text{True}}|\hat{x}_{\text{obs}})$ obtained through projection with the output PDF distribution predicted by the density network. The latter corresponds to the PDF (blue histograms in Fig. 3(c)) calculated using the network output parameters of the single-Gaussian density network. These two distributions appear to have approximately the same mean and variance around the best-fit parameter. However, unlike the distribution of the density network model calculated as exactly one Gaussian, the distribution of the true parameter is not exactly a single-Gaussian shape. This difference can be immediately resolved by using an MDN with many Gaussians, which will be shown later. As a result, the output standard deviation of the density network is shown to well approximate the CI.

Because the CI determination method using the previously described $g(x_{\text{True}}|\hat{x}_{\text{obs}})$ is a well-known method in statistics, it can be equally used for parameter estimation using a general FCP ANN. In the case of an FCP, because the output standard deviation cannot be calculated immediately, the CI can be obtained numerically by solving the integral equation in Equation (5) using $g(x_{\text{True}}|\hat{x})$. However, if we use the MDN, we obtain the CI of the parameter directly without additional analysis. The MDN method is efficient because the time required to obtain the estimated calculation results for the experimental data are similar to those of the FCP ANN model.

### 3.2 Many-Gaussian mixture density network

We now apply the multimodal mixture density model assuming several Gaussian distributions. As a result of training while increasing the number of Gaussians by a multiple of 2, the probability density distribution of the output parameters hardly changes for more than eight Gaussians. From now on, the MDN with eight Gaussians will be described. After training using an MDN with eight Gaussians, the experimental data were used as a network input to obtain a network output of eight Gaussians. To examine the distribution of multimodal PDFs obtained from the network output in detail, we first compared the PDFs obtained from the density network using one Gaussian and the MDN using eight Gaussians for the top Pt thickness parameter (Fig. 4(a)). These two PDFs have similar mean values and overall spread; however, the many-Gaussian distributions are divided into two peaks. The prediction of multiple peaks in the parameter space by the MDN for the XRR data is due to the inverse problem of X-ray scattering. That is, because the X-ray scattering intensity satisfies a functional relationship with respect to sample structure variables but does not satisfy a functional relationship inversely, several sample structures that can explain the same scattering data are predicted. However, in case of using a general FCP or a single-Gaussian density network, only one best-fit parameter is given for each parameter. Therefore, this network is not a problem if it satisfies one-to-one correspondence between input and output training data, whereas the relationship between the input and output is difficult to predict when there are two or more outputs for one input (Carmona-Loaiza & Raza, 2021; Unni *et al.*, 2020). However, in the case of an MDN that uses several Gaussian mixtures, when there are several outputs for a given input, it can be expressed in the form of a probability distribution of several peaks. Therefore, as shown in Fig. 3(a) and 3(b), the MDN predicts that each output parameter has several possible values for the experimental data given as input.

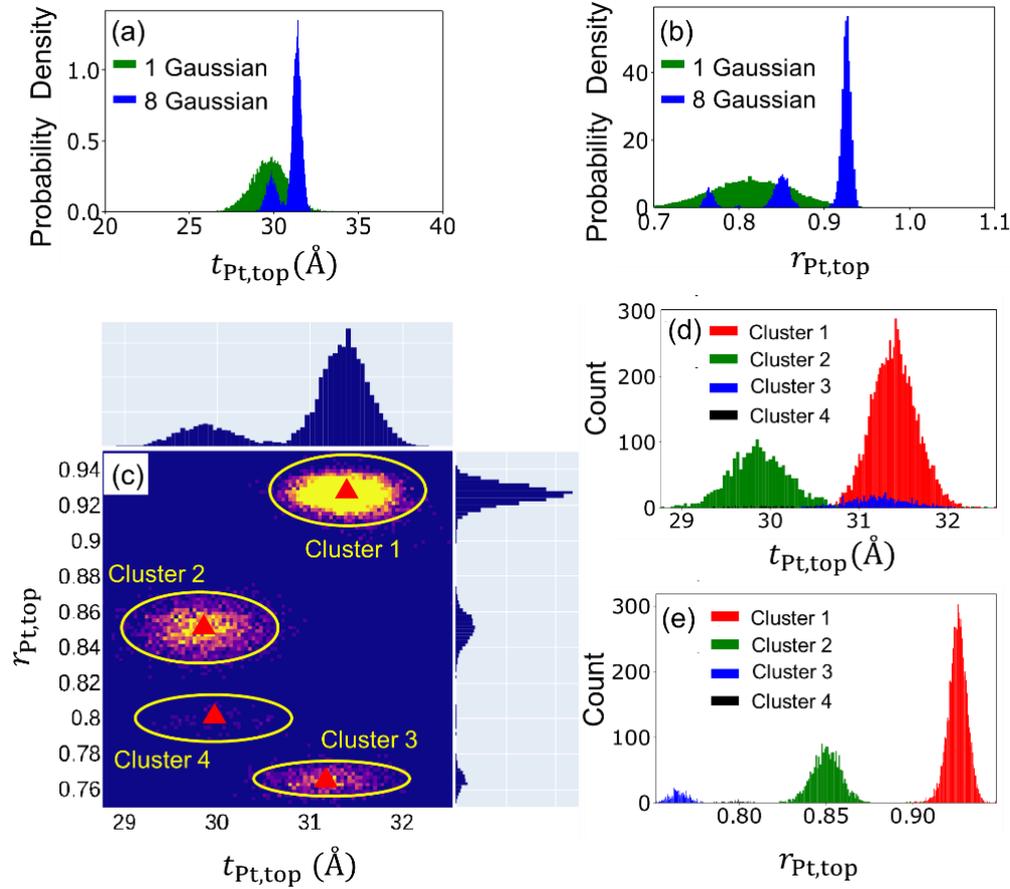

**Figure 4** (a)–(b) Probability density functions (PDFs) obtained using single-Gaussian and many-Gaussian mixture density networks. The PDFs of the thickness (a) and density (b) parameters of the top Pt layer predicted for the experimental data. The total probability of each PDF is normalized to one. (c) Two-dimensional histogram of the predicted probability distribution for the thickness and density parameters of the top Pt layer. For multidimensional histograms of all parameters, clusters are found in order of high probability using K-means clustering, which is machine learning. (d)–(e) Probability distribution for each cluster for the thickness (d) and density (e) parameters of the top Pt layer. To compare the absolute amplitudes of each cluster, the probability distribution for each cluster was not normalized.

Importantly, when sampling a parameter vector using the PDF predicted by an MDN, each parameter may appear to be intertwined rather than independent of each other. However, because the MDN theoretically ignores correlation between parameters, the Gaussian's multidimensional joint probability distribution predicted by the MDN is interpreted to have several clusters in the parameter space. Figure 4(c) shows a two-dimensional joint probability distribution for the relative density ($r_{Pt,top}$) and thickness

($t_{\text{Pt\_top}}$) of the top Pt layer, and several Gaussian combinations of the two parameters are observed to be divided into several clusters.

In the output distribution assuming a single Gaussian, because the output PDF is determined by only one peak, the best-fit parameter is obvious even in the multidimensional parameter space. However, in many-Gaussian mixture models, the number of possible parameter vectors that can be classified is difficult to determine intuitively. In particular, finding a parameter set that corresponds to the same peak in a multidimensional parameter space consisting of two or more parameters in an intuitive manner is difficult. A contour that considers the number of matching cases for all parameters can be drawn; however, quantitatively finding optimal parameters with this approach is also difficult, as is determining which peak in each parameter space corresponds to (or overlaps) which peak in another parameter space.

### 3.3 Clustering in multidimensional joint probability distribution

To determine how many Gaussian clusters are in the multidimensional joint probability distribution obtained by sampling the PDF predicted by the MDN and to find the location of the Gaussian mean of each cluster, we apply K-means clustering, an unsupervised machine learning method (Lloyd, 1982). The K-means clustering method can easily determine the number of clusters in the multidimensional space and the location of the cluster centroid. To implement K-means clustering, we used the Scikit Learn library (Pedregosa *et al*., 2011). We found a total of four clusters by carrying out an optimization by changing the number of clusters. Because the probability of the fourth cluster was small, we hereafter ignore it. The centroids of the cluster agree well with the two-dimensional Gaussian joint mean (Fig. 4(c)). Finally, these centroid values are the possible solutions that the MDN predicts for the experimental data. Figure 4(d) and 4(e) shows the clusters for the density ($r_{\text{Pt,top}}$) and thickness ($t_{\text{Pt\_top}}$) parameters of the top Pt layer, respectively. When clustering is completed, the standard deviation of the distribution for each cluster can be readily calculated.

**Table 2**  Three most probable parameter sets obtained for the experimental data using the eight-Gaussian MDN model and K-means clustering. Parameter definitions are as in Table 1.

|  | $r_{\text{Pt,top}}$ | $r_{\text{Co}}$ | $r_{\text{Pt,bot}}$ | $t_{\text{Pt,top}}$ (Å) | $t_{\text{Co}}$ (Å) |
|---|---|---|---|---|---|
| Most | $0.93 \pm 0.01$ | $0.96 \pm 0.01$ | $0.91 \pm 0.01$ | $31.4 \pm 0.3$ | $27.1 \pm 0.3$ |
| Second Most | $0.85 \pm 0.01$ | $0.96 \pm 0.01$ | $0.97 \pm 0.01$ | $29.9 \pm 0.4$ | $27.1 \pm 0.3$ |
| Third Most | $0.77 \pm 0.01$ | $0.96 \pm 0.01$ | $1.04 \pm 0.01$ | $31.2 \pm 0.3$ | $26.4 \pm 0.3$ |
|  | $t_{\text{Pt,bot}}$ (Å) | $\sigma_{\text{air/Pt}}$ (Å) | $\sigma_{\text{Pt/Co}}$ (Å) | $\sigma_{\text{Co/Pt}}$ (Å) | $\sigma_{\text{Pt/sub}}$ (Å) |
| Most | 29.7 ± 0.2 | 4.3 ± 0.1 | 5.7 ± 0.3 | 4.1 ± 0.2 | 0.0 ± 0.2 |
| Second Most | 31.2 ± 0.2 | 3.7 ± 0.1 | 4.8 ± 0.3 | 4.6 ± 0.2 | 0.1 ± 0.3 |
| Third Most | 30.6 ± 0.2 | 0.5 ± 0.2 | 2.1 ± 0.2 | 6.3 ± 0.5 | 3.5 ± 0.2 |

Table 2 shows the parameter set obtained by classifying the cluster in the aforementioned method. For the three probable solutions, most parameters show little difference; however, a notable difference appears in the roughness parameter. In the case of the highest- and second-highest-probability solutions, the roughness of the surface is greater than that of the Pt/substrate interface, whereas the roughness is reversed for the third solution. The advantage of the MDN is that it enables us to determine which solutions are more physically possible cases. Figure 5 shows the XRR curves calculated using three probable solutions. The XRR curves of the solutions with the highest and second-highest probabilities are similar to the XRR curve of the single-Gaussian solution shown in Fig. 3(a); however, their oscillation periods at high angles better explain the experimental data. This result shows that many-Gaussian MDNs can predict parameter sets more precisely than single-Gaussian MDNs. Another notable point is that the XRR curve for the solution with the third-highest probability shows the largest amplitude of XRR oscillation at high angles, which appears to better explain the experimental data. Although it is necessary to examine whether each structural variable is a physically meaningful value in order to finally determine the structural variable sets, the solution with the third-highest probability is the best in terms of whether it describes the characteristic shape of the XRR curve well. If the method of minimizing $\chi^2$ is used, then obtaining a solution sensitive to this pattern is difficult because only the solution with the highest probability is provided. By contrast, because the neural network is sensitive to the pattern of the input data, another advantage of MDN is that it provides several solutions of different features.

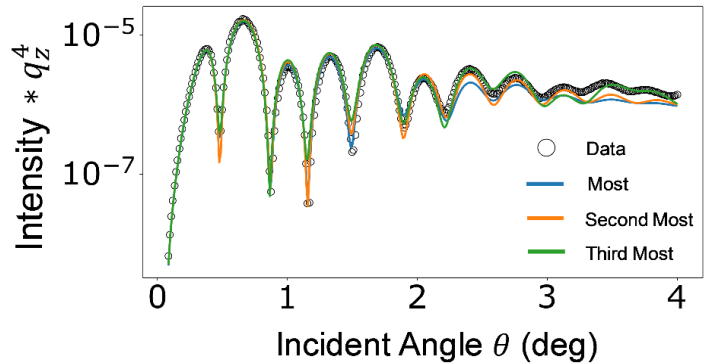

**Figure 5**  XRR curves calculated with the three most probable parameter sets obtained using an eight-Gaussian MDN.

When applying the standard least square method using the different solutions obtained from the MDN as an initial estimate, most of the regression process is completed without significantly deviating from the given initial estimate. This is because several solutions obtained by MDN each correspond to the local minima of the $\chi^2$ space. From this point of view, MDN corresponds to a neural network version of advanced regression methods for finding the global minimum in the $\chi^2$ space, for example, Bayesian analysis based on the Monte-Carlo method (McCluskev et al., 2020; Kim et al., 2021).

## 4. Conclusion

We showed that the problem of estimating the CI, which could not be obtained with the existing ANN method, was solved using the MDN method in curve fitting of XRR data. To this end, the results of the Neumann process, a conventional statistical analysis method for obtaining the CI, and the estimation results obtained using an MDN were compared. Because the MDN could well model the distribution of true parameters for the experimental data, the reliability of the estimated parameters was confirmed. This result could be used in future XRR data analyses related to the mass production of similar thin-film structures. In addition, we showed that MDN can help solve the inverse problem in X-ray scattering because it provides the PDFs instead of a single solution of the output parameters. In particular, using the K-means clustering method, possible output parameter sets from the PDFs were systematically obtained in the order of probability. The XRR data analysis method using MDN introduced here is expected to be useful for parameter estimation in other fields involving X-ray scattering or neutron reflectivity.

**Acknowledgements**   The authors gratefully acknowledge the assistance of Yong Seong Choi at the 4-ID-D beamline at the Advanced Photon Source. The authors also thank Dong-Ok Kim for his help during data acquisition.

**References**

Abadi, M., Agarwal, A., Barham, P., Brevdo, E., Chen, Z., Citro, C., Corrado, G. S., Davis, A., Dean, J., Devin, M., Ghemawat, S., Goodfellow, I., Harp, A., Irving, G., Isard, M., Jia, Y., Jozefowicz, R., Kaiser, L., Kudlur, M., Levenberg, J., Mane, D., Monga, R., Moore, S., Murray, D., Olah, C., Schuster, M., Shlens, J., Steiner, B., Sutskever, I., Talwar, K., Tucker, P., Vanhoucke, V., Vasudevan, V., Viegas, F., Vinyals, O., Warden, P., Wattenberg, M., Wicke, M., Yu, Y. & Zheng, X. (2016). *arXiv*: 1603.04467.

Behnke, O., Kröninger, K., Schott, G., Schörner-Sadenius, T. (2013). *Data Analysis in High Energy Physics : A Practical Guide to Statistical Methods*. Weinheim: WILEY-VCH.

Bishop, C. M. (1994) *Mixture density networks*. Technical Report NCRG 4288, Neural Computing Research Group, Aston University, Birmingham.

Carmona-Loaiza, J. M. & Raza, Z. (2021). *Mach. Learn.: Sci. Technol.* **2**, 025034.

Cowan, G. (1998) *Statistical Data Analysis.* Oxford: Oxford University Press.

Daillant, J. & Gibaud, A. (2009). *X-ray and Neutron Reflectivity: Principles and Applications.* Berlin, Heidelberg: Springer-Verlag.

Dillon, J. V., Langmore, I., Tran, D., Brevdo, E., Vasudevan, S., Moore, D., Patton, B., Alemi, A., Hoffman, M., Saurous, R. A. (2017). *arXiv*: 1711.10604.

Feng, Z., Hou, Q., Zheng, Y., Ren, W., Ge, J.-Y., Li, T., Cheng, C., Lu, W., Cao, S., Zhang, J. & Zhang, T. (2019). *Computational Materials Science*. **156**, 310–314.

Greco, A., Starostin, V., Karapanagiotis, C., Hinderhofer, A., Gerlach, A., Pithan, L., Liehr, S., Schreiber, F. & Kowarik, S. (2019). *J Appl Crystallogr*. **52**, 1342–1347.

Greco, A., Starostin, V., Hinderhofer, A., Gerlach, A., Skoda, M. W. A., Kowarik, S. & Schreiber, F. (2021). Mach. Learn.: Sci. Technol. 2, 045003.

Henke, B., Gullikson, E. & Davis, J. (1993). *At. Data Nucl. Data Tables*, 54, 181–342.

Ikemoto, H., Yamamoto, K., Touyama, H., Yamashita, D., Nakamura, M. & Okuda, H. (2020). *J Synchrotron Rad*. **27**, 1069–1073.

Kim, K. T., Kim, D.-O., Kee, J. Y., Seo, I., Choi, Y., Choi, J. W. & Lee, D. R. (2021). *Current Applied Physics*. S1567173921001164. DOI:10.1016/j.cap.2021.04.025.

Liu, S., Melton, C. N., Venkatakrishnan, S., Pandolfi, R. J., Freychet, G., Kumar, D., Tang, H., Hexemer, A. & Ushizima, D. M. (2019). *MRS Communications*. **9**, 586–592.

Lloyd, S. (1982). *IEEE Trans. Inform. Theory*. **28**, 129–137.

McCluskey, A. R., Cooper, J. F. K., Arnold, T. & Snow, T. (2020). Mach. Learn.: Sci. Technol. 1, 035002.

Mironov, D., Durant, J. H., Mackenzie, R. & Cooper, J. F. K. (2021). Mach. Learn.: Sci. Technol. 2, 035006.

McLachlan, G.J. and K.E. Basford, (1988) *Mixture Models : Inference and Applications to Clustering*. New York:Marcel Dekker.

Park, W. B., Chung, J., Jung, J., Sohn, K., Singh, S. P., Pyo, M., Shin, N. & Sohn, K.-S. (2017). *IUCrJ*. **4**, 486–494.

Parratt, L. G. (1954). *Phys. Rev.* **95**, 359–369.

Pedregosa, F., Varoquaux, G., Gramfort, A., Michel, V., Thirion, B., Grisel, O., Blondel, M., Prettenhofer, P., Weiss, R., Dubourg, V., Vanderplas, J., Passos, A., Cournapeau, D., Brucher, M., Perrot, M., & Duchesnay, E. (2011). J. Mach. Learn. Res., 12, 2825–2830.

Strauß, D. J., Steidl, G. & Welzel, U. (2004). *Applied Numerical Mathematics*. **48**, 223–236.

Unni, R., Yao, K. & Zheng, Y. (2020). *ACS Photonics*. **7**, 2703–2712.